# Eliminating solvents and polymers in high-performance Si anodes by gas-phase assembly of nanowire fabrics

*Moumita Rana,[+] Afshin Pendashteh,[+] Richard Schäufele, Joaquim Gispert, and Juan J. Vilatela\**

Dr. M. Rana, Dr. A. Pendashteh, R. Schäufele, Dr. J. J. Vilatela

IMDEA Materials Institute, Tecnogetafe. Eric Kandel, 2, 28906, Getafe, Madrid, Spain.

E-mail: juanjose.vilatela@imdea.org

[+] These authors contributed equally to this work.

R. Schäufele

Department of Applied Physics, Universidad Autónoma de Madrid, Cantoblanco, 28049, Madrid, Spain.

J. Gispert

Chemical Engineering and Material Science Department, IQS School of Engineering, Universitat Ramon Llull, Via Augusta 390, Barcelona, 08017, Spain.

Developing sustainable battery electrode manufacturing methods is particularly pressing for alloying-type active materials, such as silicon, which often require additional energy-intensive and solvent-based processing to reinforce them with a buffer matrix. This work introduces a new method to fabricate Si anodes as continuous, tough fabrics of arbitrary thickness, without processing solvents, polymeric binders, carbon additive, or any reinforcing matrix. The anodes consist of percolated networks of long Si nanowires directly assembled from suspension in the gas phase, where they are grown via floating catalyst chemical vapour deposition. A high Si content above 75 wt.% in a textile-like network structure leads to high-performance electrode properties. Their gravimetric capacity is 2330 mAh g$^{-1}$ at C/20 for all thicknesses produced, reaching areal capacities above 9.3 mAh cm$^{-2}$ at C/20 and 3.4 mAh cm$^{-2}$ at 1C (with 3.4 mg cm$^{-2}$). Analysis of rating data gives a high transport coefficient (6.6x10$^{-12}$ m$^2$ s$^{-1}$) due to a high out-of-plane electrical conductivity (0.6 S m$^{-1}$) and short solid-state diffusion length. Si remains a percolated network of elongated elements after extended cycling, preserving electrical conductivity and leading to a capacity retention of 80% after 100 cycles at C/5 and ~60% after 500 cycles at C/2. When integrated with NMC111 cathode, a full cell gravimetric energy density of 480 Wh kg$^{-1}$ is demonstrated.

## 1. Introduction

The rapid adoption of batteries for transport and grid storage, particularly lithium-ion batteries (LIBs), has caused a pressing need to make the materials they rely on more sustainable in terms of sourcing, manufacturing, and end-of-life. In electrode manufacture, which represents more than 25% of the energy contributions in current battery cell manufacturing,[1,2] this has translated into efforts to eliminate solvents[3] and reduce energy-intensive steps during electrode processing.[4] Sustainable manufacture is particularly challenging for emerging alloying/conversion active materials, such as Si, which often require elaborate additional processing steps to realise their high capacity for realistic electrode operation conditions.

Silicon is an abundant element, possessing about an order of magnitude higher lithium storage capacity than graphite, and is thus established as the anode for the next generation of high-energy-density LIBs. However, exploiting the high theoretical capacity of Si has been difficult due to its low electrical conductivity, the slow solid-state diffusion of Li in Si, and the fast capacity fading due to large volume change (e.g., over 300%) during consecutive Li uptake/release. Multiple approaches have been proposed to realise silicon's high capacity and reduce capacity fading,[5] but they typically introduce additional processing steps, often involving large amounts of solvents. The general strategy is to form a "buffer" matrix to compensate for the granular nature of the active material: the quasi-spherical particles are immobilised by the matrix phase and remain in contact with the internal conducting material upon repeated volumetric deformations during lithiation-delithiation cycles. One avenue has been to introduce polymer-based materials that first enable processing and then form a continuous matrix to contain the volumetric expansion of Si, for example, using cellulosic binders,[6] supramolecular polymers,[7] conducting polymers,[8] carbonised polymers,[9,10] and different fibrous polymeric networks.[11] However, most of these methods yield Si mass fractions below 60% (e.g., see literature summary in [9]) and introduce multiple additional processing steps. Similarly, Si micro/nanoparticles can be interspersed in a "matrix" of percolated high-aspect ratio nanocarbons such as carbon nanotubes[12] or graphene[13]. By providing both mechanical reinforcement and relatively high internal electrical conductivity at low mass fractions, nanocarbons can eliminate the need for polymeric binders and enable the fabrication of thick electrodes with near-theoretical capacity.[12] Despite their success, nanocarbon-reinforcing methods are inherently based on solution-processing of high aspect ratio nanocarbons, which inherently involves very dilute nanocarbon dispersions (~ 0.2 wt. %) and thus large solvent fractions.

An alternative approach is to structure electrodes by changing the shape of the active material, for instance, as silicon nanowires (Si NWs) entangled as networks in the form of freestanding solid without the need for reinforcing polymers or nanocarbons. In pioneering work, Chockla et al. synthesized a silicon nanowire fabric through a solvothermal method, followed by filtering and purification.[14] Binder-free Si-based electrodes have also been prepared from nanowires produced by acid wet-etching of monolithic Si,[15,16] reaching capacities as high as 2300 mAh g$^{-1}$ with 78% retention after 60 cycles at C/4. However, a common feature of these examples is the large consumption of organic solvents during the raw material synthetic reaction, purification, or subsequent electrode processing.

Overall, current methods to produce Si-based (and other) anodes are constrained by the use of solvents, requirement of large polymer/nanocarbon contents, and introduction of complex additional processing steps, which do not augur in favour of sustainable processing upon industrialisation. As a scalable alternative solution to overcome these limitations, this work introduces a radically new method to produce freestanding macroscopic fabrics of SiNWs through direct assembly of ultralong Si nanowires as they grow suspended in the gas phase. The SiNW fabrics have mechanical properties similar to office paper (e.g., tensile strain-to-break above 10%), can be made of arbitrary thickness, and are flexible in bending. After deposition of a thin conducting layer (also through a solvent-free process), they form electrodes with near theoretical capacity above industrially relevant thicknesses. The as-prepared anodes have a high areal capacity (>3.0 mAh cm$^{-2}$) at rates above 1C due to a high out-of-plane electrical conductivity and elimination of solid-state diffusion limitations, as well as capacity retention above state-of-the-art through preservation of the Si network upon long cycling. The anodes reported here combine high-performance properties and a gas-phase manufacturing process that is inherently scalable and can substantially reduce the environmental footprint of the next generation LIB cells.

## 2. Results and Discussion

The new electrode fabrication route is based on producing nanostructured silicon electrodes by growing Si NWs suspended in a gas stream and directly assembling them as a macroscopic material. **Figure 1**a-c shows a schematic of the process and photographs of samples before and after carbonization. The nanowires are synthesised through floating catalyst chemical vapour deposition (FCCVD), a process whereby Si NWs grow at ultra-fast rates (up to 1 μm s⁻¹) in a continuous flow reactor through thermocatalytic decomposition of silane in the presence of an aerosol of Au catalyst nanoparticles.[17] As shown in Figure 1a, a conductive layer is subsequently coated on Si NWs through pyrolysis of acetylene. After a simple transfer to a current collector, the fabric is ready to be implemented as a high Si content anode without further processing. In this fashion, the method removes altogether any need for dispersing active material or slurry preparation, thus eliminating all processing solvents. Effectively, it transforms raw material precursor (i.e., silane) into a semi-finished electrode in a single step, combining what would otherwise require several processing steps in the conventional electrode manufacturing (e.g., material fabrication, slurry making, coating, solvent recovery, drying).

As opposed to commonly done on substrates[18], growing Si NWs suspended in the gas-phase has strong implications for improving cell manufacture. It accelerates growth rate by three orders of magnitude,[17] thus increasing throughput relative to substrate-based processes. The method enables the direct organisation of nanomaterials as textiles, so far demonstrated on the meter-scale for Si NWs[17] but with the prospect of producing anodes on the kilometre scale.[19] In addition, gas-phase processes in continuous flow reactors are readily scalable[20] and used industrially for nanomaterials synthesis on the tonne scale. Furthermore, eliminating solvents and mixing processes from anode manufacture translates into a ≈ 25% energy reduction in cell manufacture[1], equivalent to reducing approximately 48.7 $kg\ eqCO_2 - kWh^{-1}$ from LIB production.[2]

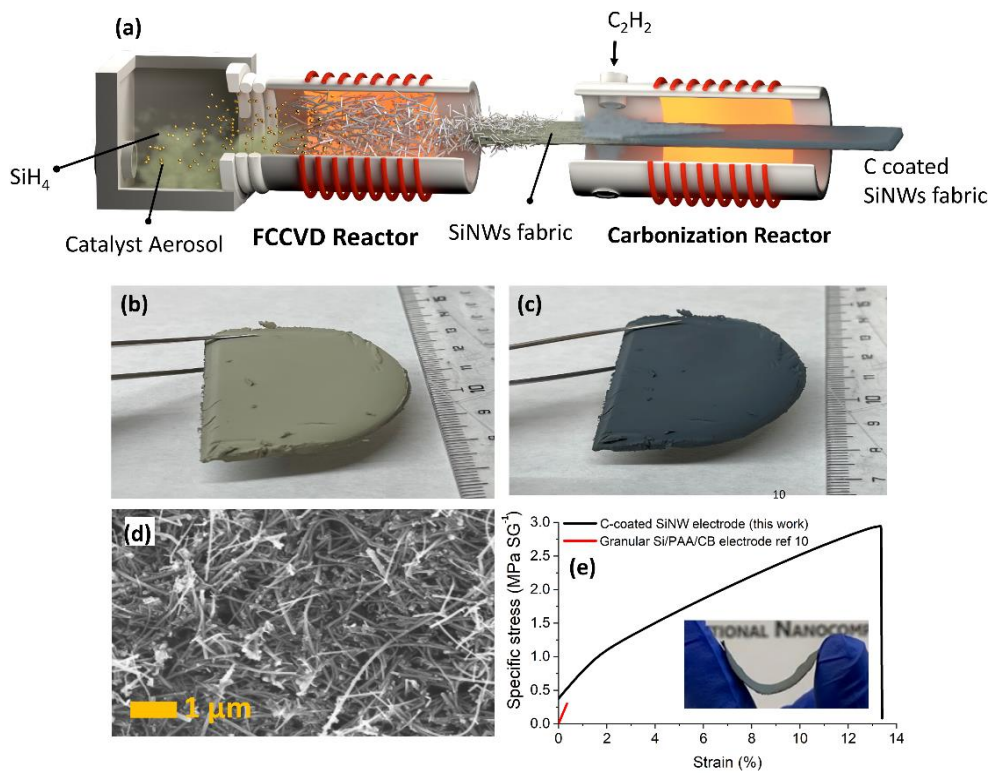

**Figure 1.** Nanostructured Si anodes assembled from gas-phase grown Si NWs followed by gas-phase C coating: (a) Schematic representation of the synthesis and assembly of silicon nanowire fabrics and carbon coating process. Photographs of freestanding samples before (b) and after C coating (c). (d) Electron micrograph showing the network structure formed by high-aspect ratio nanowires. (e) Stress-strain curve for C-coated electrodes (Si NW/C), with properties similar to office paper and flexibility in bending (inset) and over an order of magnitude above reference Si anodes without reinforcing nanofillers.[12]

Due to their high aspect ratio (~180), the Si NWs entangle and form a network (Figure 1d) capable of internal stress transfer by shear, leading to sufficient mechanical strength to form a freestanding solid material without any additive. The network structure of the electrodes makes them flexible in bending and gives them mechanical properties similar to that of other fibrous solids such as paper, with a tensile strain-to-break above 10% and specific strength up to 3MPa SG$^{-1}$ (Figure 1e). These properties are over an order of magnitude above reference granular Si anodes without reinforcing nanofillers.[12] Thanks to this mechanical robustness, the nanostructured network material can then be combined with a current collector and be handled further up to cell assembly without the need for polymer binders or other forms of mechanical reinforcement essential for conventional granular electrodes.

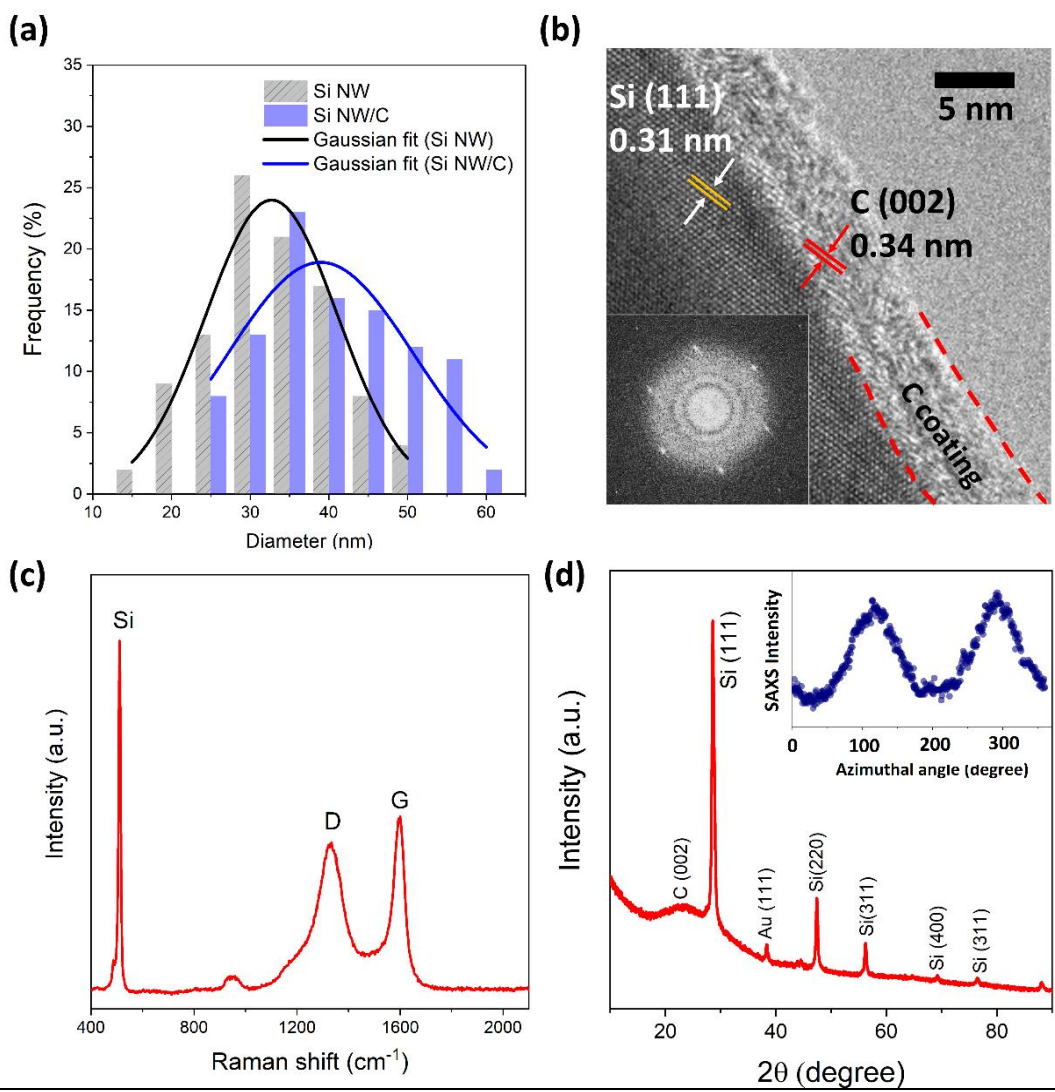

**Figure 2.** Physicochemical properties of Si NW/C fabric electrodes: (a) Histograms obtained from analysis of SEM micrographs (see Figure S1) of the pristine Si NWs network as well as the C-coated samples (Si NW/C). (b) HR-TEM micrograph of the interface between the carbon coating and the crystalline Si NW, showing preferential alignment of graphitic planes at the interface and absence of $SiO_2$. The inset displays the FFT analysis of the micrograph. (c) Representative Raman spectrum confirming the high crystallinity of the Si NWs and the moderate degree of graphitisation of the coating. (d) XRD pattern shows no evidence of $SiO_2$. The inset is the azimuthal profile from edge-on 2D SAXS measurements, indicating slight alignment of the Si NWs in the plane of the electrode.

The main morphological features of the samples before and after C-coating are observed by SEM and summarised in **Figure 2**a. The average Si NW diameter is 32 nm, which increases to 40 nm after the introduction of the carbon layer. Aiming to employ the Si NW/C fabrics as high-performance anodes for LIBs, a first imperative is to ensure that the conductive phase is uniform, sufficiently electrically conductive, and representing a relatively low mass fraction in the whole sample. For this study, the pyrolysis of acetylene at 700°C for 30 min produced a suitable balance of the three features, resulting in a high Si content electrode (above 70wt.%) with high electronic conductivity. Extensive SEM analysis (see Figure S1 and S2) of the coated samples reveals an average thickness of $3.8 \pm 1.3$ nm of the coating layer, which is uniform throughout the electrode. HRTEM reveals that the coating consists of amorphous domains and small stacks of graphitic planes with an interlayer spacing of 0.34 nm (Figure 2b).

Interestingly, it is found that the graphitic (002) basal planes nucleated on the Si NW surface are aligned parallel to Si (111) plane (Figure 2b) and thus slanted relative to the Si NW longitudinal axis. The relatively narrow G band in the Raman spectrum (Figure 2c) confirms the graphitic nature and sp$^2$ conjugation of the conductive coating. No trace of a passivating $SiO_2$ layer could be detected in the carbonised samples when analysed by XRD. This finding is in agreement with the results seen from electron microscopy, where the interface was found to consist solely of crystalline silicon and graphitic domains. The observed results indicate that the pyrolysis process results in the reduction of the surface oxides and nucleation of the solid carbon coating on the crystalline Si NW. The interfacial layer is graphitic, therefore electronically conducting, yet with the basal planes exposed to the SEI and therefore accessible for lithium diffusion to the SiNWs.

The mass fraction of the Si after introducing the conducting coating was measured through weighing the sample before and after C deposition at $73.2 \pm 1.6$ %, and confirmed via thermosgravimetric measurements analysis (see Figure S3). EDS analysis of the sample's cross-section demonstrates homogenous carbon coating of the SiNWs through-thickness of the fabric (see Figure S4). The average electrode volumetric density was 0.51 g cm$^{-3}$, equivalent to porosity of ~78%. Two-probe resistance measurements give in-plane ($\sigma_{IP}$) and out-of-plane ($\sigma_{OOP}$) electrical conductivity in the range of 1-1.5 S m$^{-1}$ and 0.1 to 0.6 S m$^{-1}$, respectively. The electrical conductivity anisotropy stems from the slight alignment of Si NWs in the electrode main plane, imparted during collection, as confirmed by edge-on SAXS measurements (see the inset of Figure 2d). Notably, the achieved $\sigma_{OOP}$ for the Si NW/C fabrics is close to the threshold of $\approx 1$ S m$^{-1}$ proposed to avoid electrical limitations in LIBs electrodes, based on semi-empirical fitting of rating data.[21] Accordingly, it can be concluded that pyrolytic reaction produces a thin (i.e., low mass fraction), uniform C layer, capable of providing sufficient electronic conductivity suitable for electrodes in LIBs. For reference, reaching $\sigma_{OOP} = 0.1$ S m$^{-1}$ in wet-processed granular electrodes with conducting nanocarbons requires volume fractions of around 10 – 20% for carbon black or graphene, or 1% of single-walled CNTs.[21]

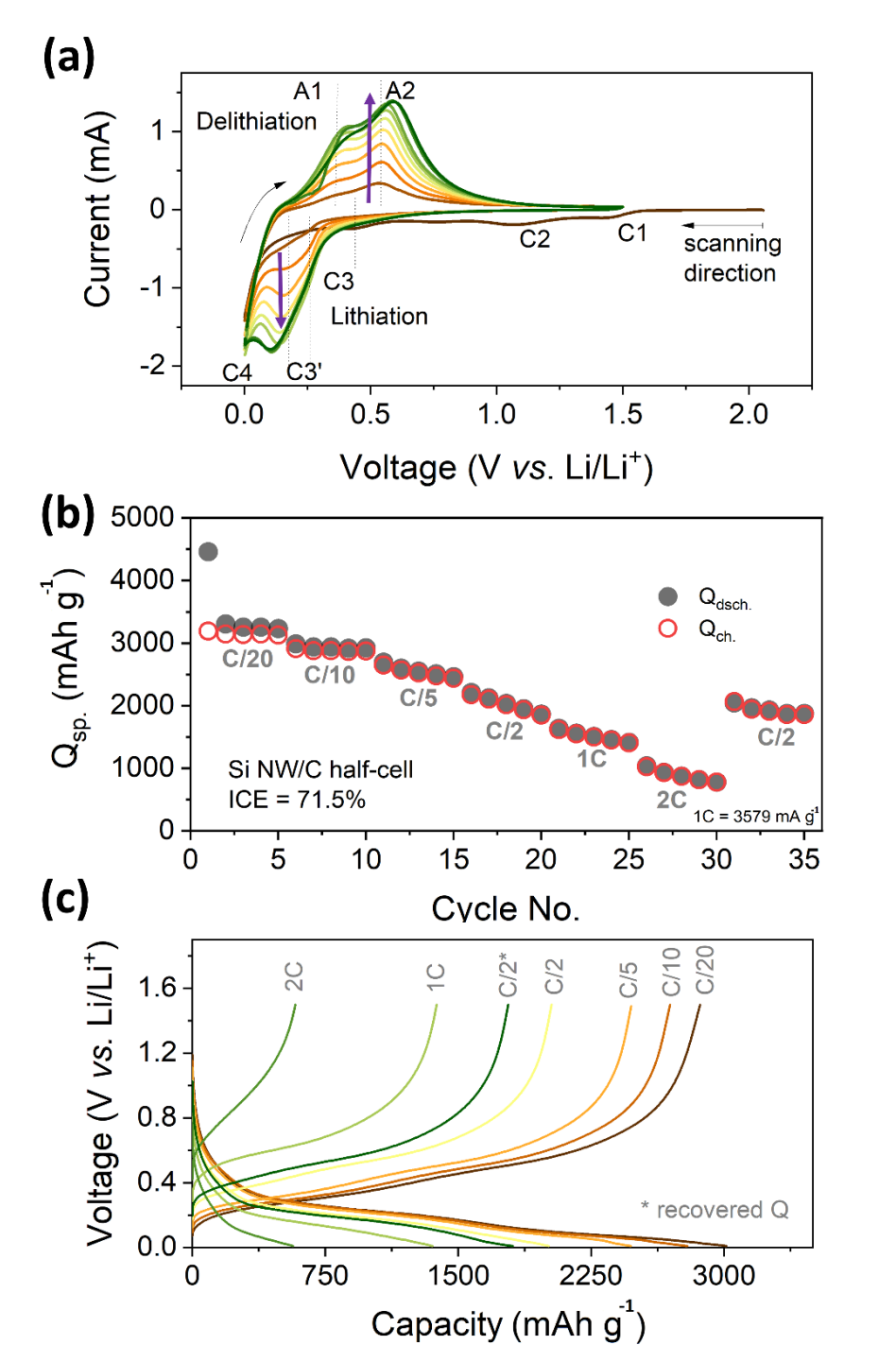

**Figure 3.** Li storage properties of the Si NW/C fabrics: (a) First ten cyclic voltammograms at a scan rate of 0.025 mV s$^{-1}$. The black arrow shows the scanning direction, and the purple arrows show the peak-intensity enhancement during cycling. (b) Rate performance of Si NW/C samples at various C rates from C/20 to 2C (1C = 3.579 A g$^{-1}$). The capacity is normalised by Si content in order to facilitate comparison with similar electrodes reported in the literature. (c) Corresponding voltage-capacity profiles. Electrode mass loading: 0.49 mg cm$^{-2}$.

The as-prepared Si NW/C fabrics were mechanically pressed onto copper foil and directly used (i.e., without any additive, binder, or use of any solvent for processing) as battery electrodes in coin cells. The electrochemical properties of the Si NW/C fabric electrodes were first evaluated using cyclic voltammetry. The first 10 consecutive cycles are shown in **Figure 3**a. During the first reduction scan, two broad cathodic peaks at potentials > 0.8 V (C1 and C2) are evident, which are ascribed to the solid electrolyte interphase (SEI) formation (i.e., irreversible reduction of the electrolyte at the electrode surface). Further sweeping the voltage towards more negative potentials, the lithiation (alloying) of Si occurs, appearing as the C3 peak, followed by the formation of $Li_{22}Si_5$ at 0.01 V.[22] In the first oxidation cycle, the peaks around 0.36 (A1) and 0.55 V (A2) can be assigned to the stepwise de-lithiation (de-alloying process) of the material through the conversion of $Li_{22}Si_5$ to $Li_{12}Si_7$ and $Li_{12}Si_7$ to amorphous Si, respectively. These results are in good agreement with those previously reported in the literature for Si-based electrodes.[22,23] In the following consecutive cycles, both cathodic and anodic peaks are shifted to higher overpotentials while the peak currents are significantly increased. This shift is attributed to electrochemical activation in the form of an increase in electrochemical surface area of the electrode upon repeated (de)lithiation process. Upon cycling, increased reaction polarisations occur due to the gradual formation and thickening of the SEI formed on the electrode surface, resulting in a longer $Li^+$ ion path and hence slower reaction kinetics.

Figure 3b shows the rate performance of the Si NW/C fabric electrodes at various c-rates ranging from C/20 ($\approx 0.179$ mA g$^{-1}$) to 2C ($\approx 7.2$ mA g$^{-1}$). The rate performance of the samples with various C content can be found in Figure S5. As it can be seen for the sample with 27 wt. % C, the first discharge capacity ($Q_{dsch.}$) is as high as 4463 mAh g$^{-1}$ (in the order of the theoretical capacity corresponding to $Li_{4.4}Si$ formation) with a high reversible charge capacity of $\sim$ 3195 mAh g$^{-1}$ (1.63 Ah cm$^{-3}$). The initial coulombic efficiency (ICE) of 71% increases up to 97% after 5 cycles at C/20. These properties may be further improved with future optimisation of the morphology and composition of the constituent SiNWs. Increasing their diameter, for example, produced an improvement of ICE to 83%, probably due to a smaller interface with the electrolyte, but at the expense of lower cycling performance (see Figure S6 and S7).

As it is seen in the Figure 3b, the capacity drops with increasing the C-rate while the CE increases up to over 99%, nevertheless still delivering a high capacity of over 780 mAh g$^{-1}$ (around 25% capacity retention, 0.38 mAh cm$^{-2}$) when the current is raised by 40-fold (i.e., at 2C). The high capacity at fast rates is partly attributed to the relatively high electrical conductivity of the electrodes due to the conductive coating. Reference samples without C coating show much lower coulombic efficiency and capacity at all rates and very rapid capacity fading (Figure S8). The conductive coating also leads to the formation of a more stable SEI, in agreement with earlier studies suggesting the role of the carbon layer in promoting the formation of a uniform SEI.[22]

More insights on the Li storage of the Si NW/C electrodes are gained from the charge-discharge voltage profiles, as shown in Figure 3c. During the first lithiation, the cell voltage decreases almost linearly down to 0.1 V, corresponding to a capacity of ~940 mAh g⁻¹. This part of the profile includes both lithiation and SEI formation, and is followed by a flat plateau around 0.1 V occurring in the course of the alloying process. Similar features are observed in the pristine Si NWs (Figure S8a), thus ruling out a significant contribution of possible irreversible reactions at the C coating. Interestingly, it can be seen in the subsequent charge-discharge profiles (Figure S8a,b) that the plateaus are shifted towards more positive potentials, corroborating that after amorphization in the first cycle, the alloying process occurs for amorphous Si. These results are in agreement with those observed from CV (Figure 3a).

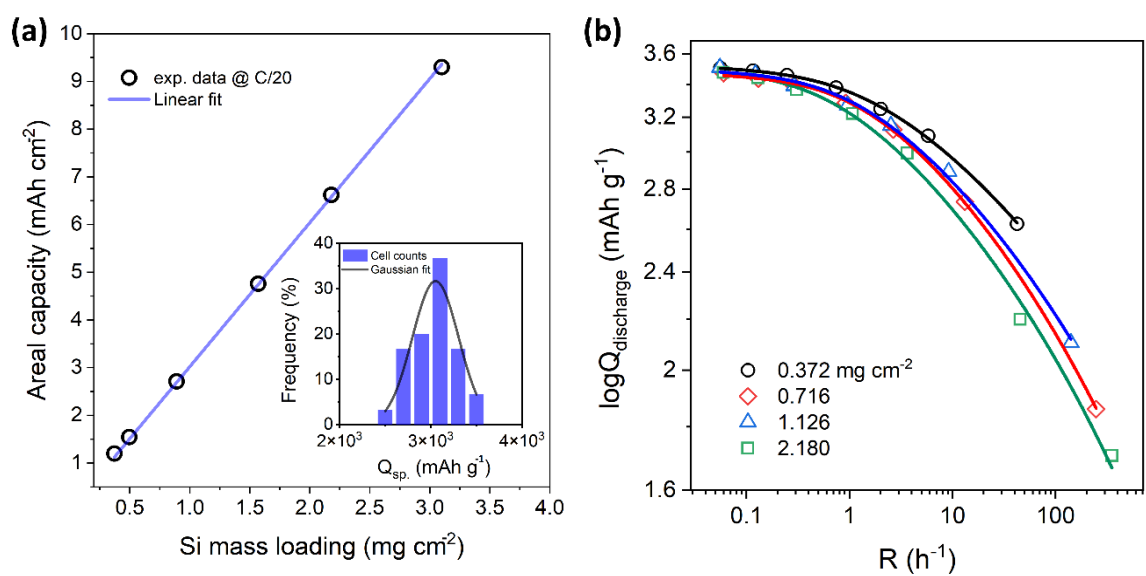

**Figure 4.** Electrochemical properties for Si NW/C anodes of increasing thickness. (a) A plot of areal capacity vs. silicon mass loading. Inset shows a histogram for the specific capacity ($Q_{sp.}$) of the cells. (b) Comparison of experimental rate performance of Si NW/C electrodes (data points) with different mass loadings (i.e. different thicknesses) along with their corresponding fittings (solid lines).

From a manufacturing perspective, 1D percolated networks become increasingly easy to manipulate as they are made thicker. An example of a > 2 mm thick sample is shown in Figure

S9. In contrast, granular electrodes crack upon drying and are limited to a critical (crack) thickness below around 175 μm. To demonstrate the capability of this new fabrication process to make thick electrodes with high areal capacity reproducibly, we fabricated binder-free, freestanding Si NW electrodes with different mass loadings ranging from 0.3 to 3.1 mg cm$^{-2}$ (with no indication of the latter being an upper limit). **Figure 4**a shows that the areal capacity (at C/20) of the Si NW/C fabrics linearly increases as the electrode mass loading (i.e., electrode thickness) increases. A specific capacity of 3041 mAh g$^{-1}$ (R$^2$ = 0.99) is obtained from linear fitting of the experimental data, confirming near-theoretical capacity accessible even for extremely thick electrodes (e.g., 300 μm). The highest areal capacity measured in these samples was 9.3 mAh cm$^{-2}$, which is amongst the highest values ever reported for binder-free Si electrodes[9,24,25] and more than sufficient to match conventional cathodes for practical full cells with a typical areal capacity of 6 mAh cm$^{-2}$.[26] Similarly, at a high charge-discharge rate of 1C, the areal capacity reaches an impressive 3.4 mAh cm$^{-2}$.

To shed light on the factors limiting capacity at high rates for thicker electrodes, we performed rating measurements on electrodes with different mass loadings, as shown in Figure 4b. For these, the rate shown is defined as $R = \frac{I_{sp.}}{(Q_{sp.})_{Exp}}$ where $I_{sp.}$ and $Q_{sp.}$ are the specific current and specific capacity at a given current, respectively. The experimental data were then analysed using a semi-empirical model recently applied to study both cathodes and anodes[12,27] and which relates the rate and specific capacity through

$$Q_{sp} = Q_M \left[ 1 - (R\tau)^n \left( 1 - e^{-(R\tau)^{-n}} \right) \right]$$

Where ɕ is the characteristic time, $Q_M$ the low-rate specific capacity, and $n$ is a rate-limiting mechanism-related parameter. As observed in Figure 4b, the fitting is in good agreement with the experimental capacity-rate data in all of the samples with various mass loadings (see Fitting parameters in Table S1). Values for $n$ span from 0.62 to 0.72, probably due to capacitive contributions, nevertheless at the low end of reported values for diffusion-limited LIB anodes (0.25 - 2).[28] The transport coefficient of the samples, $\Theta = L_E^2/\tau$, which accounts for the sample thickness ($L_E^2$), comes out as 6.6×10$^{-12}$ m$^2$ s$^{-1}$. No values of $\Theta$ have been previously reported for Si anodes, but the experimental result is in the range for other electrodes.[28]

The dominant rate-limiting process can be determined using equation: $\tau = aL_E^2 + bL_E + c$, where $a$ and $b$ include the electronic and ionic conductivity terms of the electrode and electrolyte, respectively, while $c$ represents the term for solid-state diffusion.[28] According to the fitting, the $c$ parameter was estimated to be 612 s which is over three times smaller than the value obtained for μ-Si/CNT electrode (~ 2027 s),[28] on account of the short diffusion length corresponding to the diameter of the Si NW used here. Indeed, electrochemical impedance spectroscopy measurements give a high Li$^+$ diffusion coefficient of ~$3.0 \times 10^{-12}$ in comparison with previous reports for Si nanoparticles (e.g., $10^{-13}$).[29–32] On the other hand, from the values obtained for the parameters *a* and *b*, we determine electronic conductivity limitations to become increasingly limiting only at high rates and for thick electrodes, for example, at 1C rate for electrodes of 300 μm thickness.

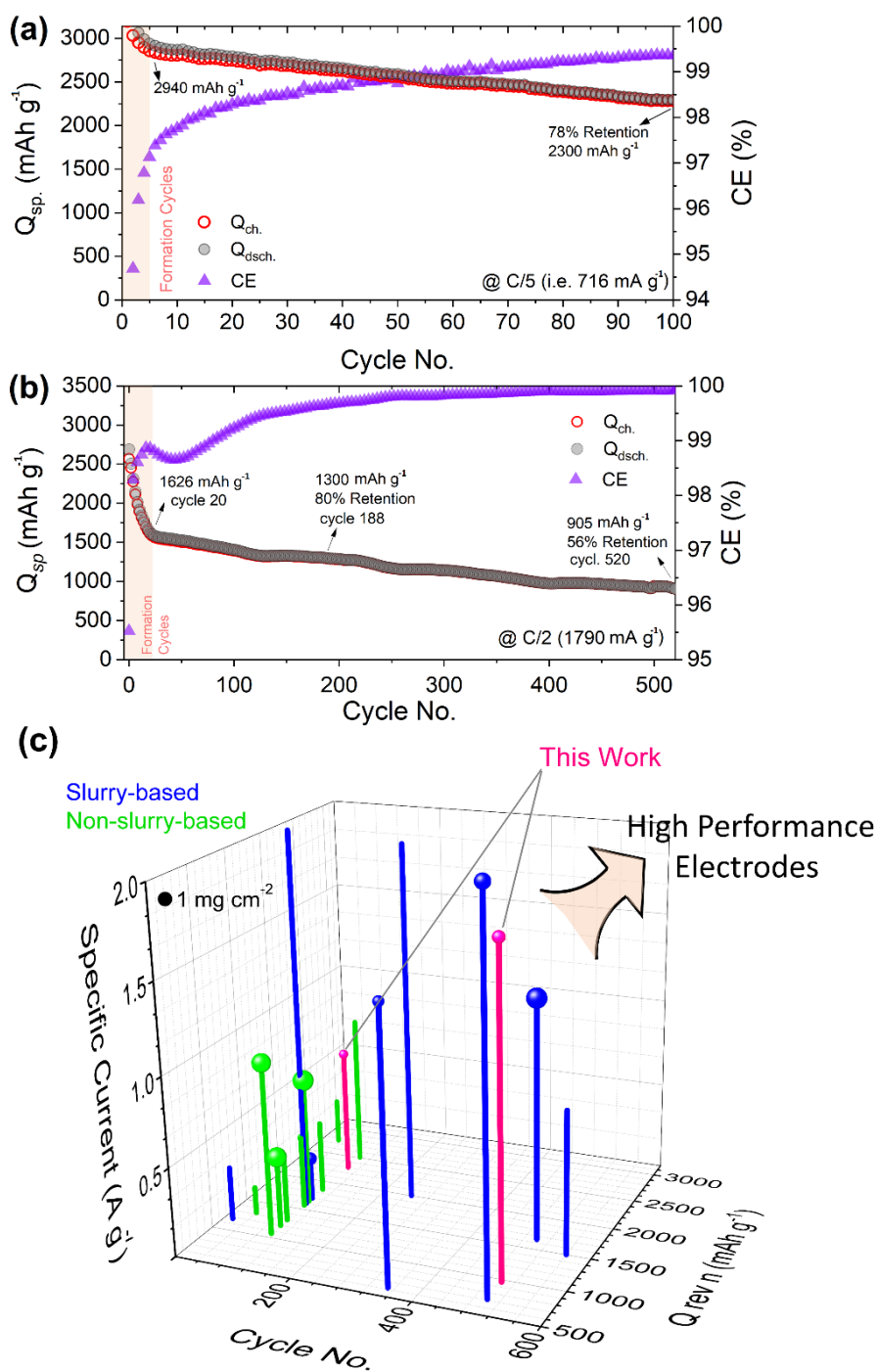

**Figure 5.** Extended galvanostatic charge-discharge cycling of Si NW/C fabric electrodes at (a) C/5 for 100 cycles and (b) at C/2 for 520 cycles as well as their corresponding coulombic efficiency (CE). (c) Comparison of the Li storage properties of previously reported binder-free silicon nanowire electrodes (slurry based and non-slurry based) after cycling, with those of this work. The size of each bar tip represents the mass loading of the Si sample.

The cycling stability of the Si NW/C fabrics was probed via consecutive galvanostatic charge-discharge cycles at various currents; the results are shown in **Figure 5**. As seen in Figure 5a, $Q_{sp}$ drops from ~ 3050 mAh g$^{-1}$ at C/5 to 2940 mAh g$^{-1}$ at the 5$^{th}$ cycle while the CE increases significantly from the ICE of 71% to over 97%. Afterwards, capacity decreases with a much more gradual slope (while the CE keeps increasing), reaching 2300 mAh g$^{-1}$ and CE > 99% after a hundred deep cycles. Given the presence of two slopes, the initial cycles before reaching ~98% CE can be considered formation cycles, and therefore capacity retention is calculated relative to that at such point. Thus defined, capacity retention at C/5 is 78% after 100 cycles and 56% (1575 mAh g$^{-1}$) after 250 cycles (Figure S10a). Figure 5b shows the cyclability of the Si NW/C electrode at a higher rate of C/2 (i.e., 1790 mA g$^{-1}$). In the first 20 cycles, the capacity fades rapidly to 1626 mA g$^{-1}$, but then the fading rate gets much smaller. This may be attributed to the fact that at higher current densities, the formation of the stable SEI requires more cycles. At C/2, after the formation cycles (i.e., 20 cycles), capacity is over 1600 mAh g$^{-1}$, comparable to that of cells subjected to rate capability tests (see Figure 3b). The SEI presumably stabilised in earlier cycles at lower current densities. The specific capacity at C/2 is 1300 mAh g$^{-1}$ (~ 80%) and 905 mAh g$^{-1}$ (~ 58%) after 188 and 520 cycles, respectively. As expected, capacity retention improves drastically for lower DoD. For 80% DoD, for example, capacity is 79.5% after 300 cycles at C/3 (Figure S10b).

These results demonstrate a high initial capacity and capacity retention in Si NW/C anodes, which is significant considering their high Si content and the absence of a reinforcing binder or nanocarbon matrix. This aspect can be appreciated by comparing their active material-normalised capacity to other literature reports. As shown in Figure 5c (and Table S2), the Si NW/C electrodes have capacity and retention at different current densities altogether at the high end of properties. They particularly stand out against the few available examples of Si NW-based anodes produced via non-slurry-based methods.

Post-mortem SEM analysis of the Si NW/C anodes after long cycling (**Figure 6** a) reveals that the overall integrity of the Si nanowires remained, with no evidence of pulverisation of the kind encountered in particulate Si-based electrodes upon consecutive charge-discharge cycles. The high aspect ratio Si NWs remain immobilised and form an interconnected network after long cycling, preserving the bulk morphology and electrical conductivity of the electrode. However, the lateral size of Si NW/C elements that make up the network increases significantly upon cycling. Through imaging of scanning electron micrographs in back scattering mode we could discriminate the inorganic Si NW/C from the SEI surrounding them, as shown in Figure 6b (and Figure S11), and therefore determine the lateral size of the Si NW/C for samples subjected to different cycles. The resulting data show a significant increase in apparent thickness relative to the original diameter of the Si NW/C (Figure 6c), which occurs predominantly in the first 30 cycles but continuous at a slower rate up to at least 350 cycles. Indeed, STM measurements of samples after long cycling (Figure 6d-e) show that repeated cycling introduce porosity in the elongated Si active material elements (Figure 6e). EIS measurements on cycled electrodes show a constant electrical resistance and increases of both charge transfer and SEI resistances (Figure 6f, g). In view of this results, we attribute the progressive capacity fading observed in these electrodes to the continuous the growth of the SEI, including possibly towards the inner surface of the SiNWs.[33]

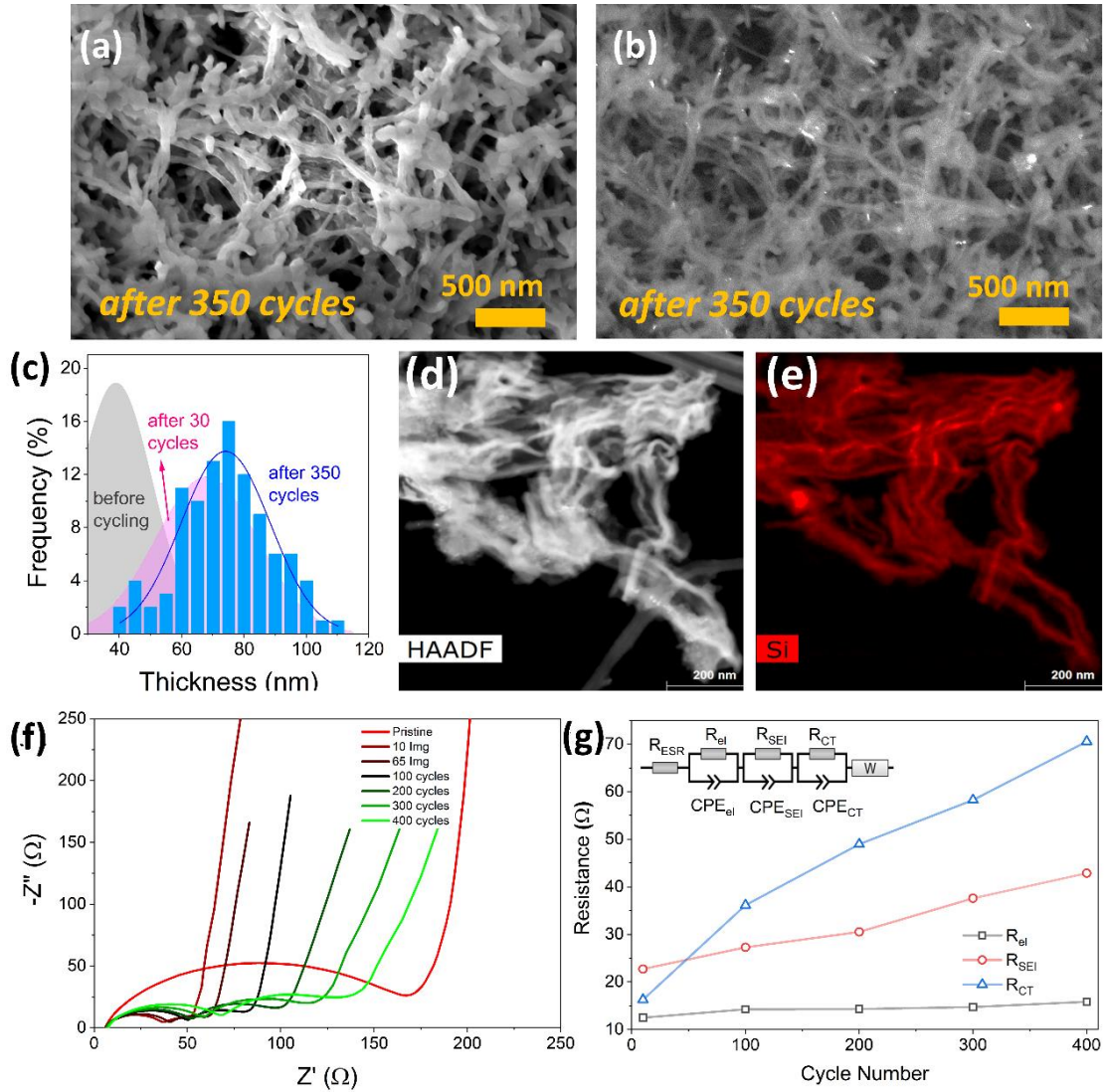

**Figure 6.** Post-mortem analysis of the Si NW/C samples after long cycling: FESEM micrograph (a) and its corresponding backscatter image (b) of the Si NW/C electrode after 350 cycles. (c) The size analysis of the nanowires after cycling in comparison with the pristine electrode. High-angle annular dark-field scanning transmission electron (HAADF-STM) micrograph (d) of the electrode after cycling and the corresponding EDS mapping for Si (e). The Nyquist plots of the Si NW/C samples at various cycle numbers (f) and various resistances in the system estimated by fitting the data (g). The inset depicts the equivalent circuit used for the fitting.

In order to probe the storage properties of the Si NW/C fabric electrodes in practice, the samples were tested in a full-cell configuration by integrating the Si NW/C electrode with a NMC111 cathode and N/P ratio of 1.02. The voltage-capacity profile of the full-cell at C/15 can be seen in **Figure 7**a. Accordingly, an initial discharge capacity of ~ 130 mAh $g_{AM}^{-1}$ at an average voltage <V> of 3.56 V was obtained, corresponding to an anode capacity of 2845 mAh $g_{Si}^{-1}$ and cathode capacity of 140 mAh $g_{NMC}^{-1}$. Figure 7b depicts the capacity retention of the full-cell over 25 cycles. Capacity gradually decreases, retaining 70% of the cell initial capacity ($\sim$ 93 mAh $g_{NMC}^{-1}$) over these cycles. The capacity fading can be attributed to the gradual capacity loss seen in the half-cell experiments, mainly due to consumption of the Li inventory in the course of SEI evolution and thickening over cycling. Further enhancement of the specific energy can be envisaged using cathodes with higher performance. Figure 7c displays calculated specific energies for integration of the Si NW/C fabric anode with currently available cathode materials (both in active material and cell level) estimated with the *CellEst* model[34]. If improvements in cell design are introduced to reduce the mass fraction of auxiliary materials to 10%, the cells with these Si NW anodes (e.g., 450 Wh kg$^{-1}$ for the cell level with LR-NMC) would surpass some performance targets set for LIB for this decade.[23]

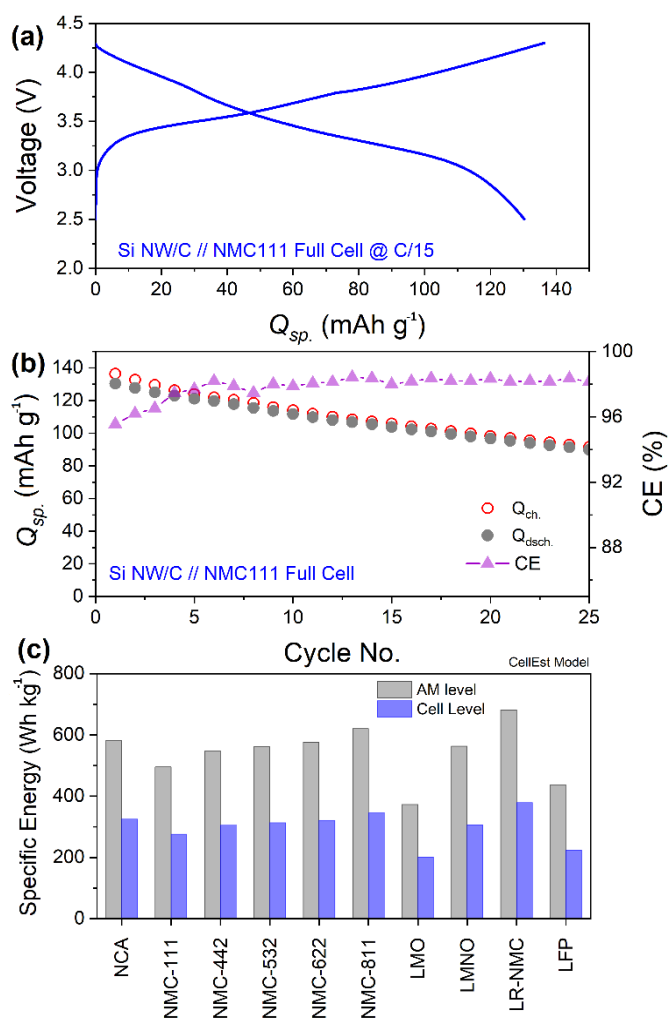

**Figure 7.** Li storage properties of the Si NW/C fabric electrodes in integration with commercial NMC111 cathode in a full-cell coin configuration. (a) Voltage-specific capacity profile of the full-cell at C/15 (1C = 150 mAh $g_{NMC}^{-1}$) in the voltage range of 2.5 to 4.3 V. (b) Cycling stability

as well as the coulombic efficiency of the Si NW/C // NMC111 full cell over 25 cycles. (c) The estimated specific energy of various cathodes in integration with Si NW/C fabric as anode, in both active material (AM) and cell levels using CellEst model.

## 3. Conclusion

This work introduces a new method to produce nanostructured Si NW fabric anodes by directly assembling nanowires as they grow floating in the gas phase, followed by deposition of a conducting C coating. The Si NW fabrics have mechanical properties similar to office paper, can be made of arbitrary thickness and are flexible in bending, which facilitates further processing and integration in battery cells. Overall, this method eliminates all solvents and polymers, in Si anode fabrication, as well as removing energy-intensive processing steps related to mixing and solvent recovery and drying. Furthermore, it simplifies the manufacturing process by transforming the raw material precursor into a continuous semi-finished electrode in one step, essentially going from gas precursor to solid electrode in a single process.

The resulting electrodes have a high Si content (over 70%) and large near-theoretical capacity at low rates (C/20), even at high areal density of 3.1 mg cm$^{-2}$ (9.3 mAh cm$^{-2}$). The high electrical conductivity and small Li diffusion length in the SiNWs provides high capacity for electrodes of thickness in the range required for existing cathodes. High capacities over 2000 mAh g$^{-1}$ could be retained after formation cycles and long cycling (e.g., 100 cycles at C/5). Extensive post-mortem analysis of cycled electrodes shows that the Si NW/C fabrics remain as a percolated network of elongated elements, which preserves electrical conductivity. However, the increase in lateral size of the Si NW/C elements upon cycling produce continuous growth of the SEI and consumption of the Li inventory. In addition to this main mechanism, others may contribute to capacity fading.[35]

Full cells have specific energy as high as 480 Wh kg$^{-1}$, relative to active materials. Coupled with introduction of cell improvements to reduce dead weight and matched with a high capacity cathodes, these anode would lead to > 400 Wh kg$^{-1}$ at cell level, which is above international performance targets for LIBs. These results show that direct assembly of nanowires into tough networks offers a potentially sustainable route to produce anodes for the next generation high energy density lithium-ion batteries.

## 4. Experimental and Methods

**Synthesis of Silicon Nanowire Fabric Electrodes**

The synthesis of silicon nanowire fabrics was performed in a vertical tubular furnace, as described in our previous work.[17] Briefly, Si NWs are synthesised via floating catalyst chemical vapour (FCCVD) at 650°C using continuous flow of an aerosol of gold nanoparticles as catalyst, $SiH_4$ as the precursor and a mixture of nitrogen and hydrogen as carrier gases. In a typical synthesis, molar fractions of gases were set at 18:31:1 ($N_2$:$H_2$:$SiH_4$) to increase reaction selectivity and ensure the samples consist of more than 95% crystalline SiNWs.[36] Due to their high aspect ratio, nanowires can be effectively collected on porous filter paper and detached to form freestanding samples, similar to non-woven fabrics or sheets. For the samples with various diameters and hence different aspect ratios the $H_2$/$SiH_4$ molar fraction ratio was changed in the range of 16 to 31. Sample areal density was adjusted through the duration of collection. The conducting phase was introduced by pyrolysis[37] of acetylene gas (5% in Ar) at 700 °C with a constant flow of 200 sccm for 30 minutes. The mass fraction of carbon in the electrode was determined gravimetrically using a 5 digit high-precision weighing balance before and after carbon coating process.

**Physiochemical Characterization**

The samples were characterized using Field Emission Scanning Electron Microscopy (FESEM, FEI Helios NanoLab 600i simultaneously acquiring images using TLD and CBS detectors), Energy Dispersive Spectroscopy (EDS), Transmission Electron Microscopy (TEM, Talos F200X FEG, 200 kV), Selected Area Electron Diffraction (SAED), Raman spectroscopy (Renishaw, using a 532 nm laser source), and 2D wide and small angle X ray scattering (WAXS and SAXS) patterns collected at the BL11-NCD-SWEET non-crystalline beamline of ALBA synchrotron light facility (Barcelona, Spain). The WAXS/SAXS patterns were collected using a microfocus spot of ~ 10-µm in diameter and a radiation wavelength of $\lambda = 1.0$ Å. The collected patterns are corrected for background scattering and obtained after radial or azimuthal integration. To determine orientation through the thickness, samples were measured with the beam parallel to the electrode plane, i.e. edge-on. Azimuthal profiles are obtained from radial integration of 2D SAXS patterns at $q = 0.035$ Å$^{-1}$. Mechanical tests on electrodes were carried out with a Favimat Textechno using a gauge length of 5 mm. Two-probe electrical resistance measurements were performed using flat copper tape contacts. Thermogravimetric analyses (TGA) of the samples were conducted using a TGA Q50 (TA instruments) with a heating rate of 5 ° min$^{-1}$.

*Electrochemical Characterization*

The performance of the carbon coated silicon nanowires as Li-ion battery anode was studied in two-electrode configuration using biologic electrochemical workstation (VMP300) and a Neware battery tester (CT-4008-5V10mA-164). Prior to cell fabrication, the C-coated Si NW fabrics were densified mechanically and dried overnight at 120 °C. Coin cells (CR 2032) were assembled with Si NW/C fabric as the working electrode, a circular lithium foil as the counter, a Whatman GF/D disc along with a PP membrane (25 um), facing the working electrode as separators, and 1 M $LiPF_6$ and 10 wt% fluoroethylene carbonate (FEC) in ethylene carbonate/diethyl carbonate (1:1 v/v, Solvionic) as electrolyte. The cells were characterized by galvanostatic cycling at the C-rates of C/20 to 2C, cyclic voltammetry at the scan rates of 0.1-10 mV $s^{-1}$ in a voltage window of 1.5-0.01 V vs. Li/Li$^+$. For the galvanostatic measurements, 1C was considered to be 3579 mAh $g^{-1}$ corresponding to the formation of $Li_{15}Si_4$ phase.[38] The cells were also characterized using electrochemical impedance spectroscopy at a frequency window of 10 mHz- 2 MHz with AC amplitude of 10 mV. Depending on the metric of interest, the specific capacity values were normalized by the mass of silicon or by the total electrode mass excluding current collector, as specified in the manuscript. For post-mortem analyses, cycled cells were disassembled inside the glove box (Ar-filled glove box with $H_2O < 0.1$ ppm and $O_2 < 0.1$ ppm). The electrodes were washed with 50 µL of anhydrous acetonitrile and dried overnight inside the glovebox before any further manipulation.

For semi-empirical analysis of the rate-dependent capacity,[21] GCD (galvanostatic charge-discharge) rate performance was determined for electrodes with different mass loadings (i.e. various thicknesses) ranging from 0.37 to 2.18 mg cm$^{-2}$. The capacity in the 5$^{th}$ cycle of each rate was considered as the stable capacity.

Positive electrode (e.g., cathode) used in the full cells was prepared using a $LiNiCoMnO_2$ (NMC111, MTI Corporation) powder as the active material, Super C65 as the carbon additive, and polyvinylidene fluoride (PVdF) binder in a mass ratio of 90:5:5 in N-methyl-2-pyrrolidone (NMP) solvent to obtain a slurry. The slurry was then caste onto cleaned Al foil using a doctor blade film applicator. The as-prepared laminate was dried at 120 °C under vacuum overnight and the punched into circle shape electrodes. The electrodes were pressed under 2 tonne cm$^{-2}$. The mass loading of the active material was 4.5 mg cm$^{-2}$.

**Supporting Information**

Supporting Information is available from the Wiley Online Library or from the author.

**Acknowledgements**

The authors acknowledge experimental support from Dr. Rebeca Marcilla from IMDEA Energía. The authors are grateful for generous financial support provided by the European Union Horizon 2020 Framework Program under grant agreement 678565 (ERC-STEM),